%% file: ICAAI-2024__main.tex
\documentclass[sigconf]{acmart}

\setcopyright{rightsretained}
\copyrightyear{2024}
\acmYear{2024}
\acmConference[ICAAI '24]{8th International Conference on Advances in Artificial Intelligence}{October 2024}{London, UK}
\acmISBN{979-8-4007-1801-4/24/10}
\acmDOI{}

\usepackage{bookmark}

\usepackage{enumitem} 
\setlist[itemize]{leftmargin=*}
\setlist{nosep,noitemsep}

\usepackage{booktabs} 
\usepackage{tabularx} 
\definecolor{lightgray}{gray}{0.97}
\let\oldtabularx\tabularx
\renewcommand*{\tabularx}{\rowcolors{1}{}{lightgray}\oldtabularx}

\usepackage{tikz} 
\usetikzlibrary{calc} 
\usetikzlibrary{shapes}
\tikzset{font={\fontsize{6pt}{8}\selectfont}}
\usepackage{adjustbox} 

\usepackage{mdframed} 
\mdfsetup{skipabove=10,skipbelow=10}

\usepackage{csquotes} 

\usepackage{hyperref} 

\begin{document}

\title[RAG: A Content Design Perspective]{Optimizing and Evaluating Enterprise Retrieval-Augmented Generation (RAG): A Content Design Perspective}

\author{Sarah Packowski}
\email{spackows@ca.ibm.com}
\affiliation{%
\institution{IBM}
\country{Canada}
}

\author{Inge Halilovic}
\email{ingeh@us.ibm.com}
\affiliation{%
\institution{IBM}
\country{United States}
}

\author{Jenifer Schlotfeldt}
\email{jschlot@us.ibm.com}
\affiliation{%
\institution{IBM}
\country{United States}
}

\author{Trish Smith}
\email{smith@ca.ibm.com}
\affiliation{%
\institution{IBM}
\country{Canada}
}

\renewcommand{\shortauthors}{Packowski, Halilovic, Schlotfeldt, and Smith}

\begin{abstract}
Retrieval-augmented generation (RAG) is a popular technique for using large language models (LLMs) to build customer-support, question-answering solutions.  
In this paper, we share our team’s practical experience building and maintaining enterprise-scale RAG solutions that answer users’ questions about our software based on product documentation.  
Our experience has not always matched the most common patterns in the RAG literature.
This paper focuses on solution strategies that are modular and model-agnostic.
For example, our experience over the past few years - using different search methods and LLMs, and many knowledge base collections - has been that simple changes to the way we create knowledge base content can have a huge impact on our RAG solutions' success. 
In this paper, we also discuss how we monitor and evaluate results.  Common RAG benchmark evaluation techniques have not been useful for evaluating responses to novel user questions, so we have found a flexible, "human in the lead" approach is required.
\end{abstract}




\maketitle

\input{ICAAI-2024_01_intro}

\input{ICAAI-2024_02_related}
\input{ICAAI-2024_03_ux}

\input{ICAAI-2024_04_accessibility}

\input{ICAAI-2024_05_architecture}

\input{ICAAI-2024_07_optimize_content}

\input{ICAAI-2024_08_strategy}

\input{ICAAI-2024_09_guidelines}

\input{ICAAI-2024_10_eval}

\input{ICAAI-2024_11_scale}

\input{ICAAI-2024_12_thanks}

\bibliographystyle{ACM-Reference-Format}
\bibliography{ICAAI-2024_bib}

\end{document}

%% file: ICAAI-2024_01_intro.tex
\section{Introduction}

Retrieval-augmented generation (RAG) is an effective way to use large language models (LLMs) to answer questions while avoiding hallucinations and factual inaccuracy\cite{towar, ragged, certified}.
Basic RAG is simple: 1) search a knowledge base for relevant content; 2) compose a prompt grounded in the retrieved content; and 3) prompt an LLM to generate output.  
For the retrieval step, one approach dominates the literature: 1) segment content text into chunks; 2) index vectorized chunks for search in a vector database; and 3) when generating answers, ground prompts in a subset of retrieved chunks\cite{survey1}. 
Our RAG solutions don't always use vector databases for search.

Wikipedia has long been influenced by and had an influence on scientific research \cite{shaped, wikipedia-impact}. With respect to RAG, Wikipedia is a dominant source of knowledge base content for training data and benchmarks, including: 
2WikiMultiHopQA, AmbigQA, ASQA, 
DART, FEVER, 
HotpotQA, KILT, MuSiQue, 
Natural Questions, NoMIRACL, PopQA, 
SQuAD, StrategyQA, SuperGLUE, 
TriviaQA, WikiAsp, 
WikiBio, WikiEval, 
and Wizard of Wikipedia\cite{
2wiki-multihop-qa, ambigqa, asqa,
dart, fever, 
hotpotqa, kilt, musique, 
natq, nomiracl, popqa, 
squad, strategyqa, superglue,
triviaqa, wikiasp, 
wikibio, ragas, wow}.
The knowledge base for our team's RAG solutions is our own product documentation, which is structured differently from Wikipedia articles.  

Using common benchmarks to test your RAG implementation involves these steps: 1) index the given knowledge base content in your retriever component; 2) prompt your solution to answer the given questions; and 3) compare generated answers to expected answers, using methods such as exact match, cosine similarity, BLEU, ROUGE, METEOR, BertScore, or using LLMs as judges\cite{eval-survey}.
Those evaluation metrics have not been useful for evaluating our RAG results for novel questions from real users.

In this paper, we share our experience building enterprise-scale RAG solutions, with a focus on three aspects:
\begin{itemize}
\item \textbf{RAG implementation} -- 
Our solutions are modular. Our retriever and generative components are closed boxes, accessed through APIs, with a limited ability to fine-tune.
\item \textbf{Knowledge base content} -- 
We are able to improve results by optimizing the knowledge base content itself.  We developed content strategy and writing guidelines for RAG.
\item \textbf{Evaluating results} -- 
We test our RAG solutions with real user questions before making the solutions available to external users. We evaluate run-time answers after the solutions are launched.
\end{itemize}

\noindent Supporting material for this paper is available on GitHub.\footnote{\url{https://github.com/spackows/ICAAI-2024_RAG-CD}}

%% file: ICAAI-2024_02_related.tex
\section{Related work}


Many techniques have been proposed for improving upon the basic RAG method.  Advanced RAG techniques include: augment knowledge base content (with metadata or knowledge graphs, for example); fine-tune the embedding model, the retriever, or the generative model (or all of them); rewrite or expand the query; use multiple knowledge bases (diverse in their format and content) and then route queries; evaluate, re-rank, filter, and post-process retrieved chunks; generate multiple outputs to choose the best one; or iteratively refine results\cite{trag, retriever, blended, raft, end2end, gar, rqrag, erag, crag, filter-context, active-rag}.
Little attention has been paid to optimizing the knowledge base content itself.

RAG solution builders must consider the structure of their knowledge base content when converting that content to text before indexing it for search.  Much work has been done exploring the best way to read PDF documents, capture meaning from HTML or Markdown elements, interpret images, and reflect relational information in tables and lists\cite{pdf,tables,nougat,telcorag}.

The structure of knowledge base content must also be considered when segmenting the content into chunks.  Chunking too small risks splitting information across multiple chunks.  Chunking too large risks including irrelevant information in a given chunk.  Choosing a chunk size depends on multiple factors, including the profile of the knowledge base content\cite{seven,telcorag}.  
One way to include a complete, self-contained idea or explanation in each chunk is to chunk content not based on size, but at the chapter or section level\cite{hiqa}.

The dominant search method in early RAG literature has been vector embeddings.  Now, combining multiple search strategies (including traditional ones\footnote{\url{https://applied-llms.org/\#dont-forget-keyword-search-use-it-as-a-baseline-and-in-hybrid-search}}) is increasing\cite{keyword-search,noembeddings}.


When evaluating results from a deployed RAG solution, multiple authors have acknowledged manual work is required\cite{ragfusion,inspectorraget,eval-survey}.
To assist with human evaluation, \cite{factscore} and \cite{factool} propose identifying facts in questions, retrieved knowledge base content, and generated answers to confirm facts agree.
ARES\cite{ares} and RAGAS\cite{ragas} validate question-context relevance, context-answer faithfulness, and question-answer relevance.
When evaluating their RAG solutions, \cite{seven} identify seven points of failure: 1) missing content; 2) search failure; 3) context window limitations; 4) poor answer generated by the LLM; 5) incorrect output format; 6) vague answers; and 7) incomplete answers.  We can confirm seeing our solutions encounter those same failure points too.

%% file: ICAAI-2024_03_ux.tex
\section{RAG implementation}

\pdfbookmark[subsection]{User interface}{User interface}
\subsection*{User interface}

\begin{figure}[!htbp]
\centering
\includegraphics[width=\columnwidth]{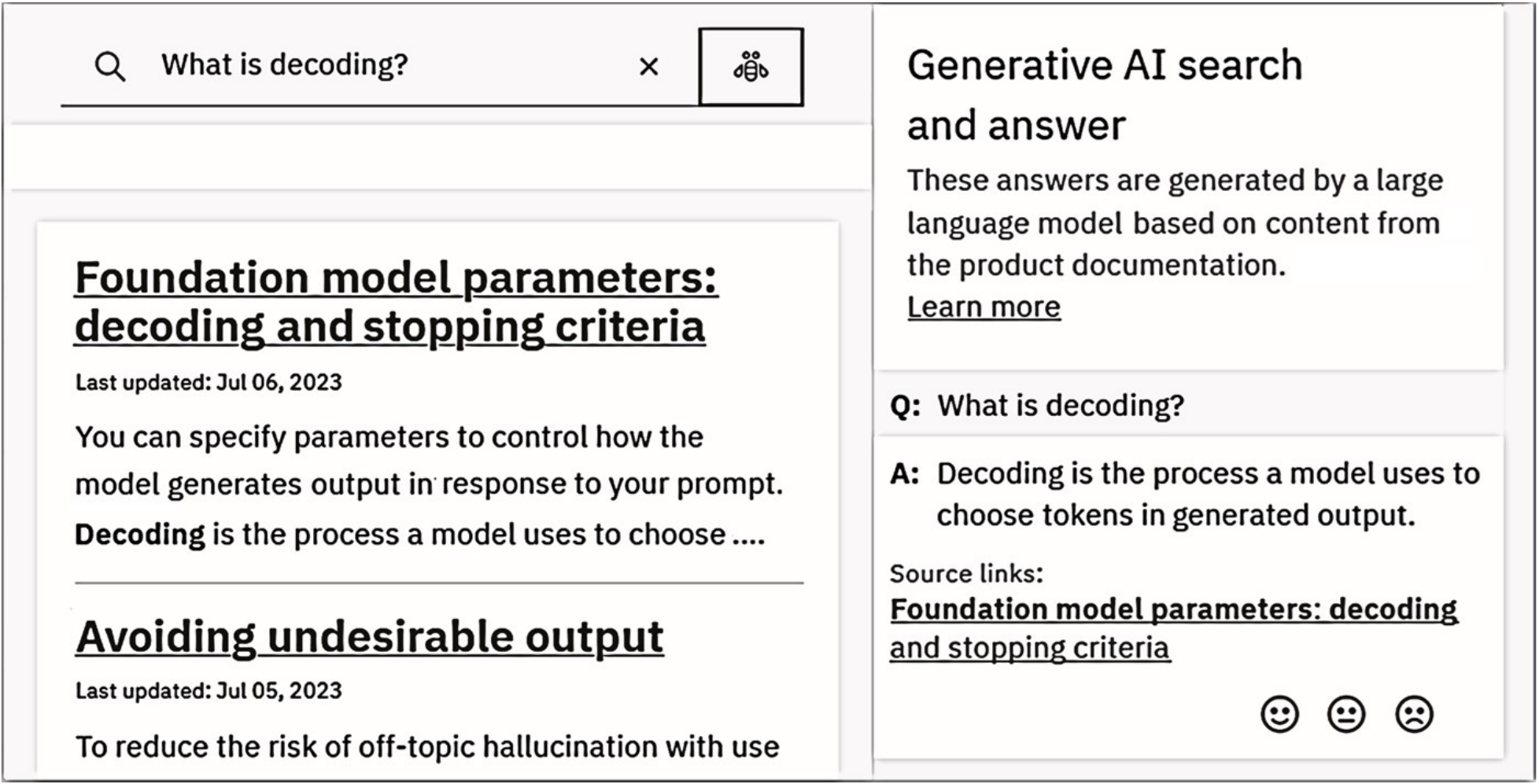}
\caption{Search-and-answer user interface}
\Description{Screen capture of a search-and-answer user interface in online software documentation.  A search bar shows a question asked by a user, a central area shows regular search results, and a side bar shows a short answer generated by an LLM as well as a link to a documentation topic in which the answer is grounded.}
\label{ux}
\end{figure}

Fig. \ref{ux} shows the interface of a simple RAG solution deployed on the search page of product documentation.  A discussion of key user experience design aspects follows.

\textbf{Search enhancement} -- 
When a search query is in the form of a natural language question, the usual search results are returned and a brief answer is generated by an LLM.  Readers are accustomed to using the search bar to look for information, so there is no new interface to discover and learn how to use.

\textbf{Not a chatbot} -- 
Previous dialog turns are not included in the LLM prompt for context.  We chose to deploy a simple solution to get experience and feedback quickly.  Also, our team is interested to explore non-chatbot LLM interfaces.

\textbf{Shaping user behavior} -- 
The search bar is a single line input, so it is awkward to type a complex question there.  This friction nudges users to keep their questions concise.  Fewer than 6\% of questions submitted to the solution are much longer than the search input or require multi-hop reasoning.  We plan to study the impact of the restricted input on that behavior.

\textbf{Transparency and explainability} -- 
Links to content in which an answer is grounded are always provided.  Also, terms that significantly impacted the way the solution generated the answer are highlighted in bold.  When we review solution logs, we see the highlighting helps users know which terms to change (or remove) in their question to get a different answer.

\begin{figure}[!htbp]
\begin{minipage}[t]{0.48\columnwidth}
  \includegraphics[width=\linewidth]{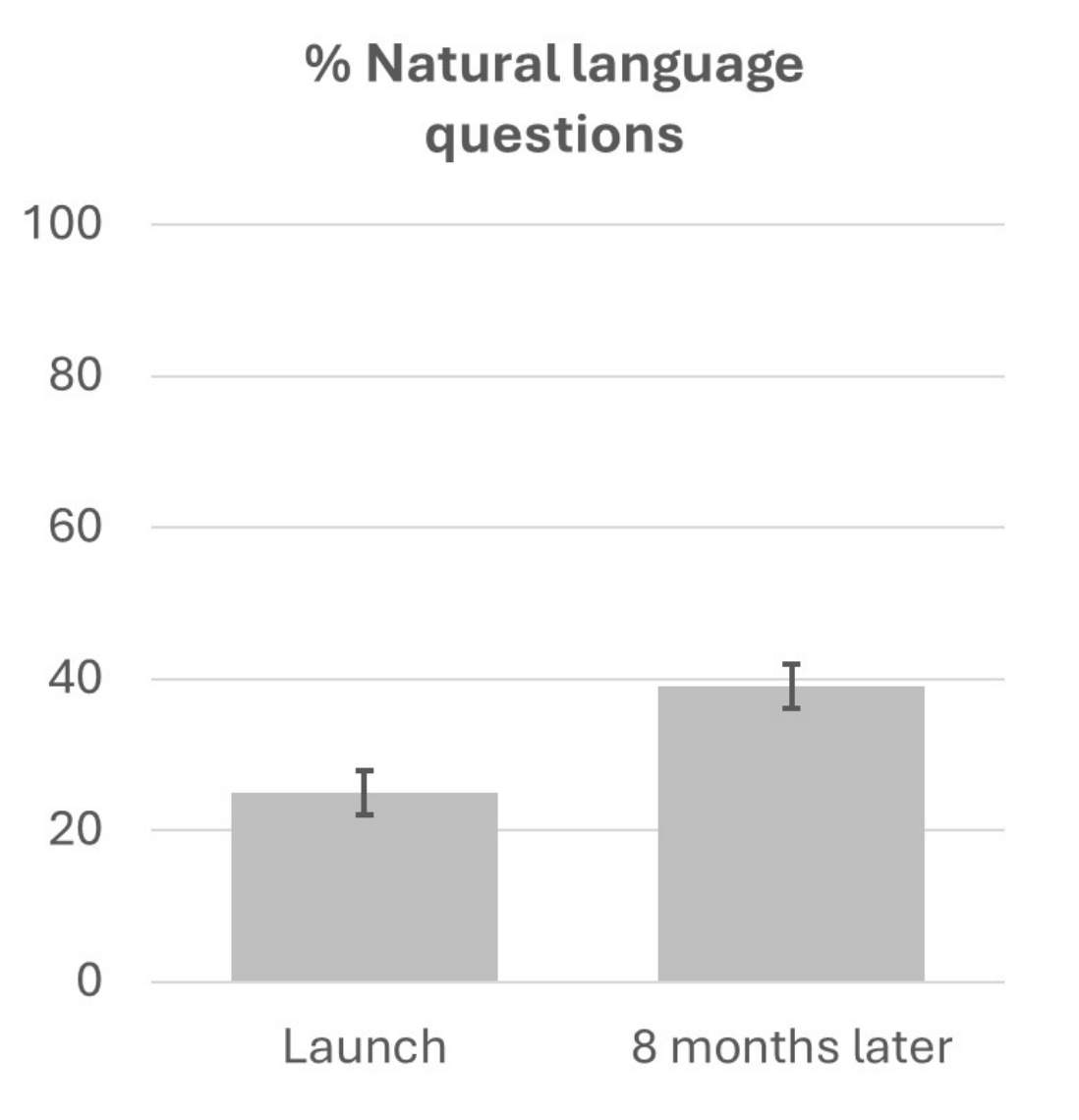}
  \caption{}
  \Description{Chart showing how the percent of input posed as a natural language question has increased since the RAG solution was launched.  The first bar shows 25 percent of input was natural language questions when the RAG solution was launched.  The second bar shows 39 percent of input was natural language questions 8 months later.}
  \label{nl-queries}
\end{minipage}\hfill 
\begin{minipage}[t]{0.48\columnwidth}
  \includegraphics[width=\linewidth]{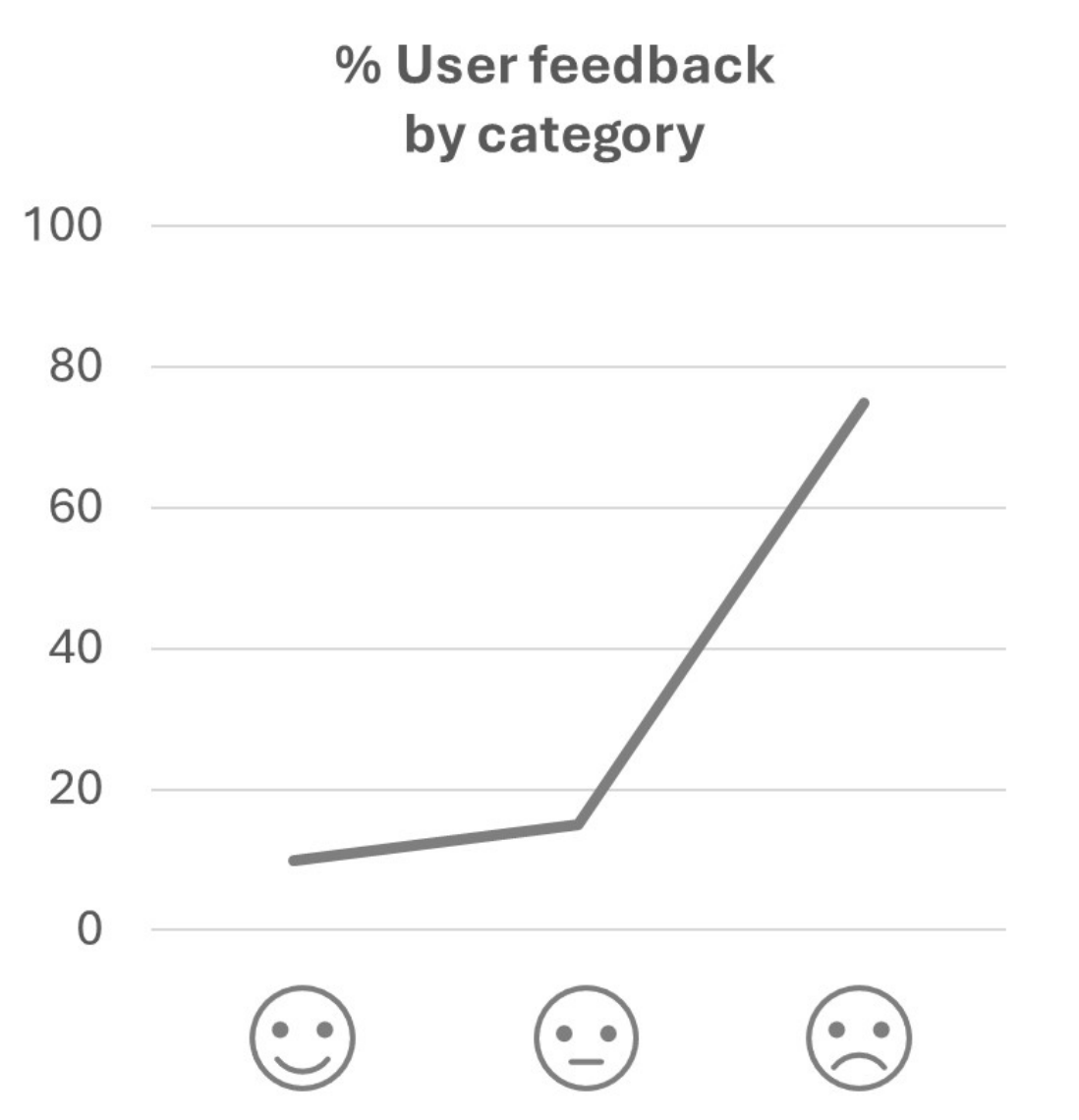}
  \caption{}
  \Description{Chart showing how user feedback is distributed amongst the feedback categories.  A line chart shows 10 percent of feedback is positive, 15 percent is neutral, and 75 percent of feedback is negative.}
  \label{user-feedback}
\end{minipage}
\end{figure}

\textbf{Growth of natural language questions} -- 
Over time, the percent of queries submitted in the search bar that are expressed as a natural language question (versus a keyword search) has increased from 25\% to 39\%.  (Fig. \ref{nl-queries})

\textbf{User feedback} -- 
We worried users wouldn't give feedback if it required multiple clicks, choosing from difficult-to-interpret categories, or typing explanations.  But we also worried simple \enquote{thumbs-up} or \enquote{thumbs down} feedback wouldn't be fine-grained enough to analyze the impact of iterative solution improvements. So, we chose a 1-click interface with three options: \enquote{helpful}, \enquote{somewhat helpful}, and \enquote{unhelpful}.  We found that users give feedback less than 1\% of the time, and mostly for unhelpful answers. (Fig. \ref{user-feedback})

%% file: ICAAI-2024_04_accessibility.tex
\textbf{Accessibility} -- 
Accessibility guidelines for static content have existed for some time.\footnote{\url{https://www.w3.org/WAI/standards-guidelines/wcag}}  And tools for testing HTML pages are readily available.
But making LLM-driven natural language interfaces accessible has many open questions:
\begin{itemize}
\item Will generated text be clear for all reading levels?
\item Can you bookmark a chat if you have memory challenges?
\item Will generated images be clear if you see colors differently?
\item Can you easily navigate generated output using a keyboard?
\item Will generated video have captions and scene descriptions?
\end{itemize}

%% file: ICAAI-2024_05_architecture.tex
\pdfbookmark[subsection]{Solution architecture}{Solution architecture}
\subsection*{Solution architecture}

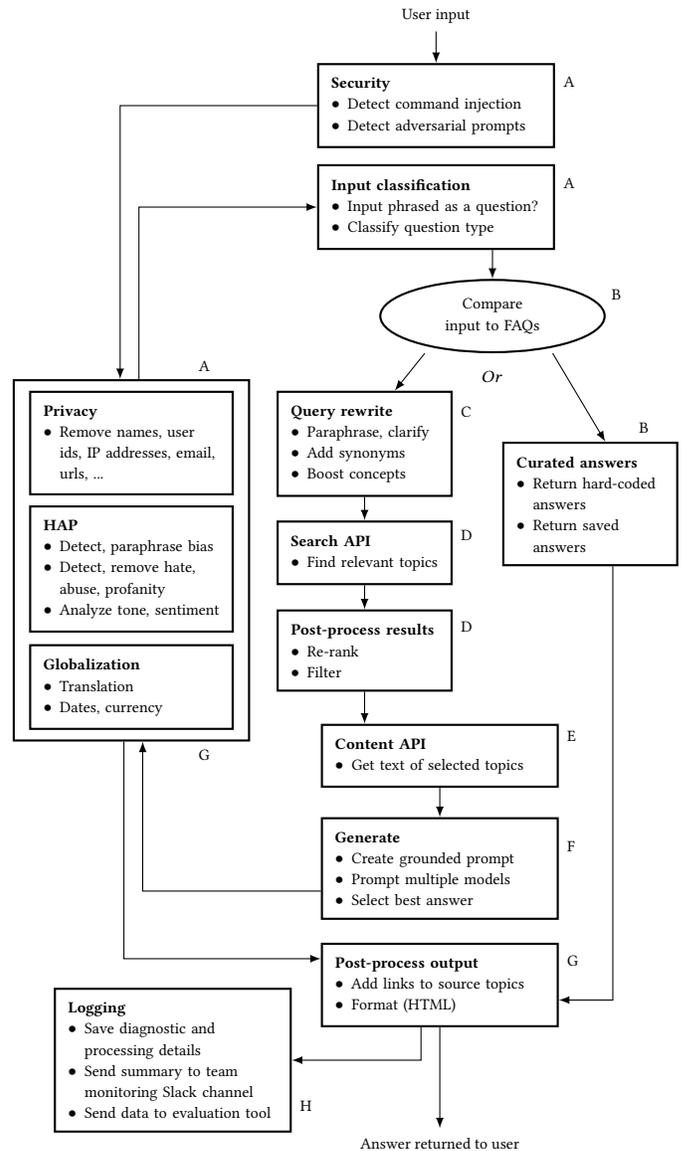
\begin{figure}[!htbp]
\centering
\input{ICAAI-2024_06_archdiag-2}
\caption{RAG solution diagram}
\Description{Architecture diagram of RAG solution components: 1) User input is filtered for security; 2) personal information is removed; 3) the query is translated to English; 4) hate, abuse, and profanity (HAP) paraphrased; 5) input is classified; 6) frequently asked questions are answered with curated answers; 7) novel questions are rewritten for clarity and to improve search; 8) search API is called; 9) search results are post-processed; 10) a prompt grounded in the retrieved topics is sent to an LLM; 11) the generated answer is post processed to remove personal information, paraphrase HAP, and translate to the original input language; 12) results are logged.}
\label{fig-arch-diag}
\end{figure}

Fig. ~\ref{fig-arch-diag} shows components of the RAG solution mentioned above.  The knowledge base is product documentation made up of \enquote{topics}, using the Darwin Information Typing Architecture (DITA) paradigm.\footnote{\url{https://dita-lang.org/1.3/dita/archspec/base/introduction-to-dita}}  A discussion of key aspects follows.

\textbf{(A) Pre-processing user input} --
Malicious input, such as JavaScript injection and adversarial prompts, is rejected; personal information removed; bias as well as hate, abuse, and profanity (HAP) paraphrased.  Input is translated to English and  classified to determine if it is a question 
and to detect the question type, such as \enquote{what-is}, \enquote{how-to}, or \enquote{troubleshooting}. Only unharmful questions move on through the solution -- in English.

\textbf{(B) Frequently asked question (FAQs)} -- 
If a user's question matches a sensitive FAQ -- related to legal terms, for example -- we return a hard-coded answer.  For other FAQs, we curate previously generated answers evaluated as useful.
Before returning a curated answer, we confirm the grounding topics have not been updated, because that could change the answer.  If the topics have been updated or deleted, the question is 
handled like a novel question.

\textbf{(C) Augmenting the question} -- 
If the user's question does not match an FAQ, the question is further processed to improve search performance: ambiguous questions rewritten; jargon replaced with in-domain terms; synonyms added.

\textbf{(D) Search as a closed box} -- 
We call a search API that returns a ranked list of topics relevant to our query.  Some search results might be re-ranked or filtered by our RAG solution.  A separate team manages the search API we use.  They maintain the API and automatically index our documentation continuously.

\textbf{(E) Whole topics instead of chunks} -- 
Once we have a list of relevant topics, we extract the complete text of those topics to ground our prompt.  Sometimes called \enquote{small2big} or \enquote{parent document retrieval}, this strategy works well for us because our topics are optimized for RAG. They are short, complete, accurate, and up-to-date. Our text-extraction component takes advantage of the reliable structure of our topics.  For example, we have writing guidelines requiring tables to be fairly simple.  Knowing this, we convert tables to row-wise and column-wise lists of lists to retain row-column relationships without worrying about complex tables.

\textbf{(F) Simple prompts} -- 
Our solution isn't a dialog, so we don't need to maintain chat history. Because the user interface encourages simple questions, because we clean, clarify, and augment questions, because topics are optimized for RAG, and because we want concise answers that are faithful to the topics, the LLM has only one job: rewrite content from the grounding topics in a succinct answer.  We prompt multiple models and choose the best answer.

\textbf{(G) Post-processing output} -- Generated output is processed like the input: personal information, bias, and HAP generated by the LLM is removed or paraphrased.  The answer is translated to the language of the original user question, links to grounding topics are added, and the whole output is marked up in HTML.

\textbf{(H) Logging} -- Because LLMs can generate problematic output, our team monitors solution activity in real time by sending detailed logging to a team Slack\footnote{\url{https://slack.com}} channel.  We also send details to an evaluation tool discussed later in this paper.

%% file: ICAAI-2024_06_archdiag-2.tex
\newenvironment{tightitemize}
{ \begin{itemize}[nosep,noitemsep,leftmargin=*,labelsep=0.5em,topsep=0pt] }
{ \end{itemize} } 

\newcommand{\userinput}{%
User input
}

\newcommand{\security}{%
\textbf{Security}
\begin{tightitemize}
\item Detect command injection
\item Detect adversarial prompts
\end{tightitemize}
}

\newcommand{\compare}{%
Compare\linebreak input to FAQs
}

\newcommand{\curated}{%
\textbf{Curated answers}
\begin{tightitemize}
\item Return hard-coded answers
\item Return saved answers
\end{tightitemize}
}

\newcommand{\queryrewrite}{%
\textbf{Query rewrite}
\begin{tightitemize}
\item Paraphrase, clarify
\item Add synonyms
\item Boost concepts
\end{tightitemize}
}

\newcommand{\search}{%
\textbf{Search API}
\begin{tightitemize}
\item Find relevant topics
\end{tightitemize}
}

\newcommand{\postprocesssearch}{%
\textbf{Post-process results}
\begin{tightitemize}
\item Re-rank
\item Filter
\end{tightitemize}
}

\newcommand{\retrieve}{%
\textbf{Content API}
\begin{tightitemize}
\item Get text of selected topics
\end{tightitemize}
}

\newcommand{\generate}{%
\textbf{Generate}
\begin{tightitemize}
\item Create grounded prompt
\item Prompt multiple models
\item Select best answer
\end{tightitemize}
}

\newcommand{\postprocessllm}{%
\textbf{Post-process output}
\begin{tightitemize}
\item Add links to source topics
\item Format (HTML)
\end{tightitemize}
}

\newcommand{\returnansewr}{%
Answer returned to user
}

\newcommand{\privacy}{%
\textbf{Privacy}
\begin{tightitemize}
\item Remove names, user ids, IP addresses, email, urls, ...
\end{tightitemize}
}

\newcommand{\hap}{%
\textbf{HAP}
\begin{tightitemize}
\item Detect, paraphrase bias
\item Detect, remove hate, abuse, profanity
\item Analyze tone, sentiment
\end{tightitemize}
}

\newcommand{\globalization}{%
\textbf{Globalization}
\begin{tightitemize}
\item Translation
\item Dates, currency
\end{tightitemize}
}

\newcommand{\logging}{%
\textbf{Logging}
\begin{tightitemize}
\item Save diagnostic and processing details
\item Send summary to team monitoring Slack channel
\item Send data to evaluation tool
\end{tightitemize}
}

\newcommand{\classify}{%
\textbf{Input classification}
\begin{tightitemize}
\item Input phrased as a question?
\item Classify question type
\end{tightitemize}
}

\newsavebox{\hapbox}
\savebox{\hapbox}{%
    \begin{tikzpicture}[%
        level distance=15mm,
        block/.style ={
            rectangle, 
            draw=black, 
            thick, 
            fill=white,
            text width=7.4em, 
            text ragged, 
            inner sep=5pt
        }
    ]
    \node (n11) [block] {\privacy}
        child { node (n12) [block,yshift=-5] {\hap} edge from parent[draw=none]
            child { node (n13) [block,yshift=-2] {\globalization} edge from parent[draw=none] }
        };
    \end{tikzpicture}%
}

\begin{tikzpicture}[%
    auto, 
    scale=1.0, 
    every node/.style={transform shape},
    sibling distance=65mm, 
    level distance=15mm, 
    edge from parent/.style={draw,-latex}, 
    block/.style ={
        rectangle, 
        draw=black, 
        thick, 
        fill=white,
        text width=8.8em, 
        text ragged, 
        inner sep=5pt
    },
    block_narrow/.style ={
        rectangle, 
        draw=black, 
        thick, 
        fill=white,
        text width=6.2em, 
        text ragged, 
        inner sep=5pt
    },
    block_xnarrow/.style ={
        rectangle, 
        draw=black, 
        thick, 
        fill=white,
        text width=5em, 
        text ragged, 
        inner sep=6pt
    },
    block_noborder/.style ={
        rectangle, 
        draw=none, 
        thick, 
        fill=none,
        text centered, 
        minimum height=1em
    },
    oval/.style ={
        ellipse,
        draw=black, 
        thick, 
        text width=6em, 
        text centered, 
        minimum height=2.25em
    },
    container/.style ={
        rectangle, 
        draw=black, 
        thick, 
        fill=white,
        text width=9em, 
        text centered, 
        inner sep=4pt
    }
]
\node (n0) [block_noborder] {\userinput}
    child { node (n1) [block,yshift=3mm,label=5:{A}] {\security}
        child { node (n16) [block,yshift=1.5mm,label=5:{A}] {\classify} edge from parent[draw=none]
            child { node (n14) [container,xshift=-8mm,yshift=-32mm,label=72:{A},label=288:{G}] {\usebox{\hapbox}} edge from parent[draw=none] }
            child { node (n2) [oval,xshift=-25mm,yshift=0.5mm,label=4.5:{B}] {\compare}  edge from parent[draw=none]
                child { node (n4) [block_narrow,xshift=48mm,yshift=-2mm,label=12:{C}] {\queryrewrite} edge from parent[draw=none] 
                    child { node (n5) [block_narrow,yshift=.5mm,label=3:{D}] {\search} 
                        child { node (n6) [block_narrow,yshift=2mm,label=7:{D}] {\postprocesssearch}
                            child { node (n7) [block,xshift=10mm,yshift=1mm,label=3:{E}] {\retrieve} edge from parent[draw=none]
                                child { node (n8) [block,yshift=0mm,label=4:{F}] {\generate} 
                                    child { node (n9) [block,yshift=-0.5mm,label=5:{G}] {\postprocessllm} edge from parent[draw=none] 
                                        child { node (n15) [block,xshift=-3mm,yshift=5mm,label=345:{H}] {\logging} edge from parent[draw=none] }
                                        child { node (n10) [block_noborder,xshift=-32.5mm,yshift=-6mm] {\returnansewr} }
                                    }
                                }
                            }
                        }
                    }
                }
                child { node (or) [xshift=0mm,yshift=7mm] {$Or$} edge from parent[draw=none] }
                child { node (n3) [block_narrow,xshift=-52mm,yshift=-10mm,label=57:{B}] {\curated} edge from parent[draw=none] }
                    }
            }
        };
\draw [->,-latex] (n1.west) -| ([xshift=-1.5mm]n14.north);
\draw [->,-latex] ([xshift=1mm]n14.north) |- (n16.west);
\draw [->,-latex] ([xshift=7.5mm]n16.south) to ([xshift=0mm]n2.north);
\draw [->,-latex] ([xshift=-9mm]n2.south) to ([xshift=4mm]n4.north);
\draw [->,-latex] ([xshift=0mm]n6.south) to ([xshift=-10mm]n7.north);
\draw [->,-latex] ([xshift=8mm]n2.south) to ([xshift=2mm]n3.north);
\draw [->,-latex] ([xshift=3mm]n3.south) |- ([yshift=-2mm]n9.east);
\draw [->,-latex] ([yshift=-3mm]n8.west) -| ([xshift=1.5mm]n14.south);
\draw [->,-latex] ([xshift=-1mm]n14.south) |- ([yshift=3.5mm]n9.west);
\draw [->,-latex] ([xshift=-2.5mm]n9.south) |- (n15.east);
\end{tikzpicture}

%% file: ICAAI-2024_07_optimize_content.tex
\section{Knowledge base content}

Our team uses search and LLM APIs that we have limited ability to adjust.  What we can control is our knowledge base content.  Our content must be easy to search, navigate, and consume by all readers, including people not working in their first language and people using tools like screen readers.  Now, we also want our content to work well for RAG solutions.

\begin{mdframed}[linecolor=gray]

\begin{center}
\textbf{Content rewriting experiment}
\end{center}

We built a simple RAG solution to answer 189 questions about \textit{the Earth} from the Natural Questions benchmark\cite{natq}.
Initially, our solution did not answer all questions correctly.

One question we failed to answer is: \enquote{what is the pre-industrial level of co2 on earth?}  The correct answer is: \enquote{280 ppm.} But our solution responded with: \enquote{180 ppm.}

Text from the relevant article used to answer that question follows.  The underlined text is the only edit needed for the RAG solution to answer correctly:

\begin{mdframed}[linecolor=gray,backgroundcolor=blue!10!gray!10]
Over the past 400,000 years, CO2 concentrations have shown several cycles of variation from about 180 parts per million during the deep glaciations of the Holocene and Pleistocene to 280 parts per million during the interglacial periods \underline{until the pre-industrial era}.
\end{mdframed} 

Minor edits like this increased success to 100\%.  Complete code and edits are available on GitHub.\footnote{\url{https://github.com/spackows/ICAAI-2024_RAG-CD}}

\end{mdframed} 

For many RAG projects that use legacy knowledge base content, rewriting that content isn't feasible.  However, for a RAG solution that is to be built 6 months from now or a year from now, the knowledge base content might not yet exist.  Some estimate that more than 250,000 websites are created every day.\footnote{\url{https://www.forbes.com/advisor/business/software/website-statistics}}  On Wikipedia, more than 400 articles are being added every day.\footnote{\url{https://en.wikipedia.org/wiki/Wikipedia:Size_of_Wikipedia}}  For our teams,  products that will be released next year don't have any documentation yet.  When creating new content, it makes sense to optimize it for RAG solutions.

%% file: ICAAI-2024_08_strategy.tex

\pdfbookmark[subsection]{Content strategy}{Content strategy}
\subsection*{Content strategy for RAG}

Testing RAG solutions before making them available to users might seem difficult due to a lack of test questions that reflect what real users will ask\cite{seven}.  However, when creating documentation for a new product or feature, content designers have always researched what questions users are likely to ask:
\begin{itemize}
\item We run internal workshops with teammates and observe where participants get stuck and what questions they ask.
\item We read internal communities where teammates ask questions as they use internal releases of upcoming features.
\item We review external forums where users are asking questions about similar functionality in other products.
\item When features are in early, limited release, we collaborate with sales, pre-sales technical support, and customer advocates to find out what questions early users have.
\end{itemize}

These are all ways to collect questions that better represent what real users will ask than any questions we might guess ourselves.  Optimizing content to be used in RAG solutions requires paying more attention to what questions that content must answer. A new quality metric will be: How well does a given topic answer anticipated user questions?

\begin{mdframed}[linecolor=gray]

\begin{center}
\textbf{Testing topics}
\end{center}

Imagine you have a list of real user questions about credentials and you have a draft topic about credentials.  How could you verify the topic answers those questions?

You could prompt an LLM to answer the user questions grounded in the draft topic, then verify the answers. Or you could prompt an LLM to generate questions answered by the draft topic, then automatically compare the generated questions to the real questions. Hypothetical draft topic:

\begin{mdframed}[linecolor=gray,backgroundcolor=blue!10!gray!10]
Credentials are the user ID and password for authenticating with the service.  Credentials are important, they prevent others using your service instance.
\end{mdframed} 

\noindent Hypothetical generated questions:

\begin{itemize}
\item What are credentials?
\item Why are credentials important?
\item What do credentials prevent?
\end{itemize}

The topic seems helpful and those seem like questions people might ask.  But what if real user questions are: \enquote{Where do I find my credentials?}, \enquote{How do I get my credentials?}, and \enquote{Where can I look up my credentials?}  The generated questions don't match those, which means a RAG solution using that topic won't work for real users.

\end{mdframed}

%% file: ICAAI-2024_09_guidelines.tex

\pdfbookmark[subsection]{Content guidelines}{Content guidelines}
\subsection*{Content guidelines for RAG}



As we monitored our RAG solution results and then experimented with rewriting content, we identified patterns that led to better results.  We asked writers from other teams to test these patterns with their content too.  From this testing, we created guidelines to help writers optimize content for RAG:

\begin{itemize}
\item \textbf{Simplify complex tables} --
Tables that have spanned cells or lack column headings are difficult for LLMs to interpret. 
\item \textbf{Explain graphics in text} --
Explaining graphics clarifies ambiguities and avoids the need for an image-to-text model.
\item \textbf{Add summaries to tutorials or long procedures} --
LLMs struggle with long tutorials or procedures (because of getting \enquote{lost in the middle} or context window limitations.)  Adding a summary is an easy way to improve results.
\item \textbf{Clearly introduce lists} --
LLMs can better use content in lists when there is a clear lead-in sentence before the list.
\item \textbf{Simplify nested content} --
Meaning can be lost by the LLM when content has multiple levels of nesting (steps with sub-steps that have option lists, for example.) Avoiding multi-level nesting improves results.
\end{itemize}

%% file: ICAAI-2024_10_eval.tex
\section{Evaluating RAG results}

\begin{figure}[!htbp]
\centering
\includegraphics[width=\columnwidth]{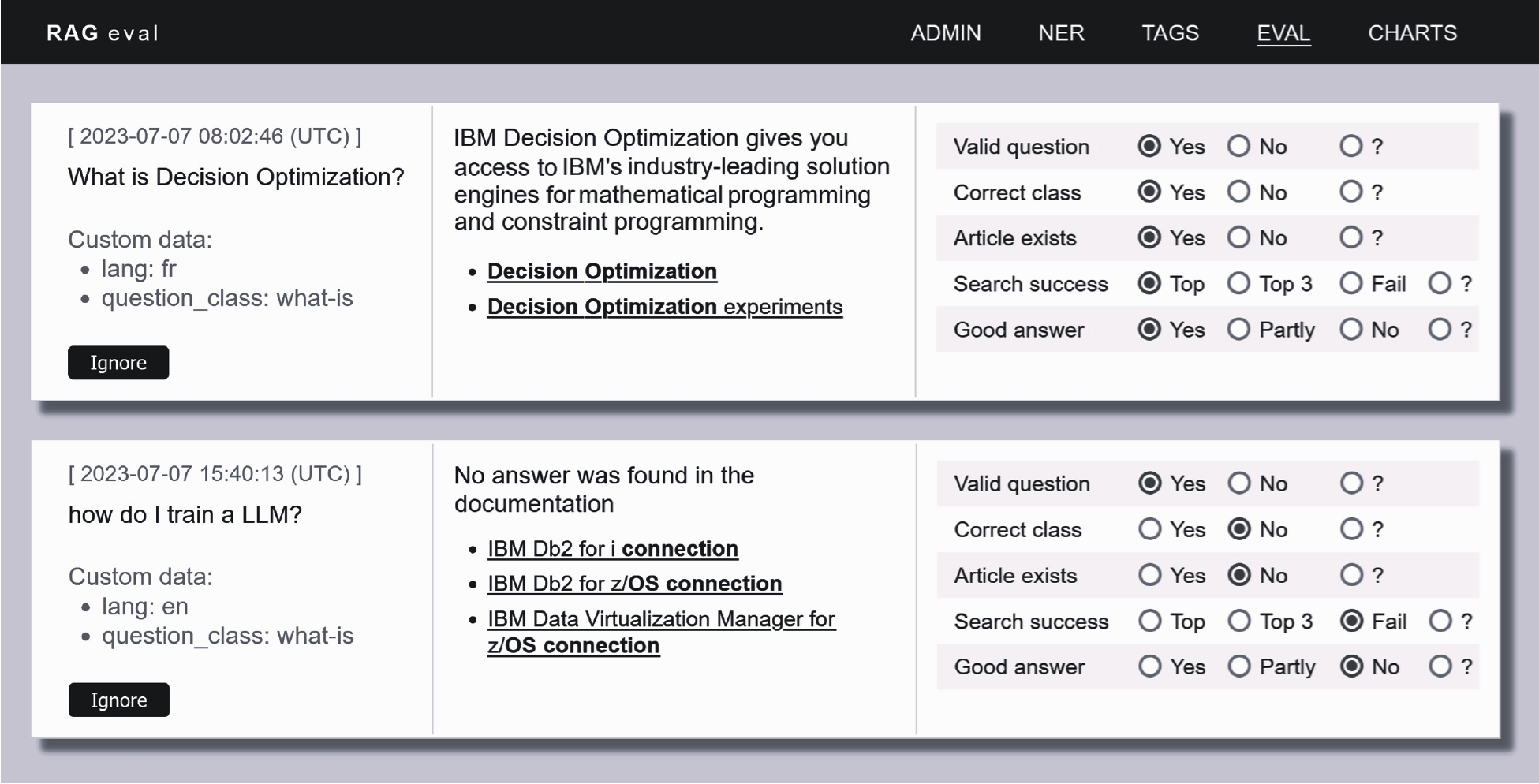}
\caption{RAG evaluation web app}
\Description{Screen capture of a web app for evaluating RAG results.  The graphical interface enables evaluators to scroll up and down to see results, filter restults based on different criteria, and check radio buttons to perform the evaluation.}
\label{rag-eval}
\end{figure}

We created a web app that streamlines the task of manually reviewing and evaluating answers our RAG solutions return to users.  Fig. \ref{rag-eval} shows the evaluation page of the app.  On the left is the user's question as well as some metadata, such as the language in which the question was submitted and the question classification.  In the middle is the answer that was returned to the user, complete with generated answer text and links to relevant topics.  On the right is a list of criteria:
\begin{itemize}
\item \textit{Valid question} - Should we be able to answer this question?
\item \textit{Correct class} - Was the question classified correctly?
\item \textit{Article exists} - Is there a topic to answer the question?
\item \textit{Search success} - Did search find the relevant topics?
\item \textit{Good answer} - Is the answer accurate, complete, helpful?
\end{itemize}

As we evaluate results, annotate key terms, and tag results, we end up naturally creating fine-tuning data sets, custom NLP dictionaries, and training data for classifiers.  Our evaluation tool takes a \enquote{human in the lead} approach: AI learns from the data our manual work naturally creates so it can automatically perform some evaluations, entity identification, and classification.  

\begin{figure}[!htbp]
\centering
\includegraphics[width=\columnwidth]{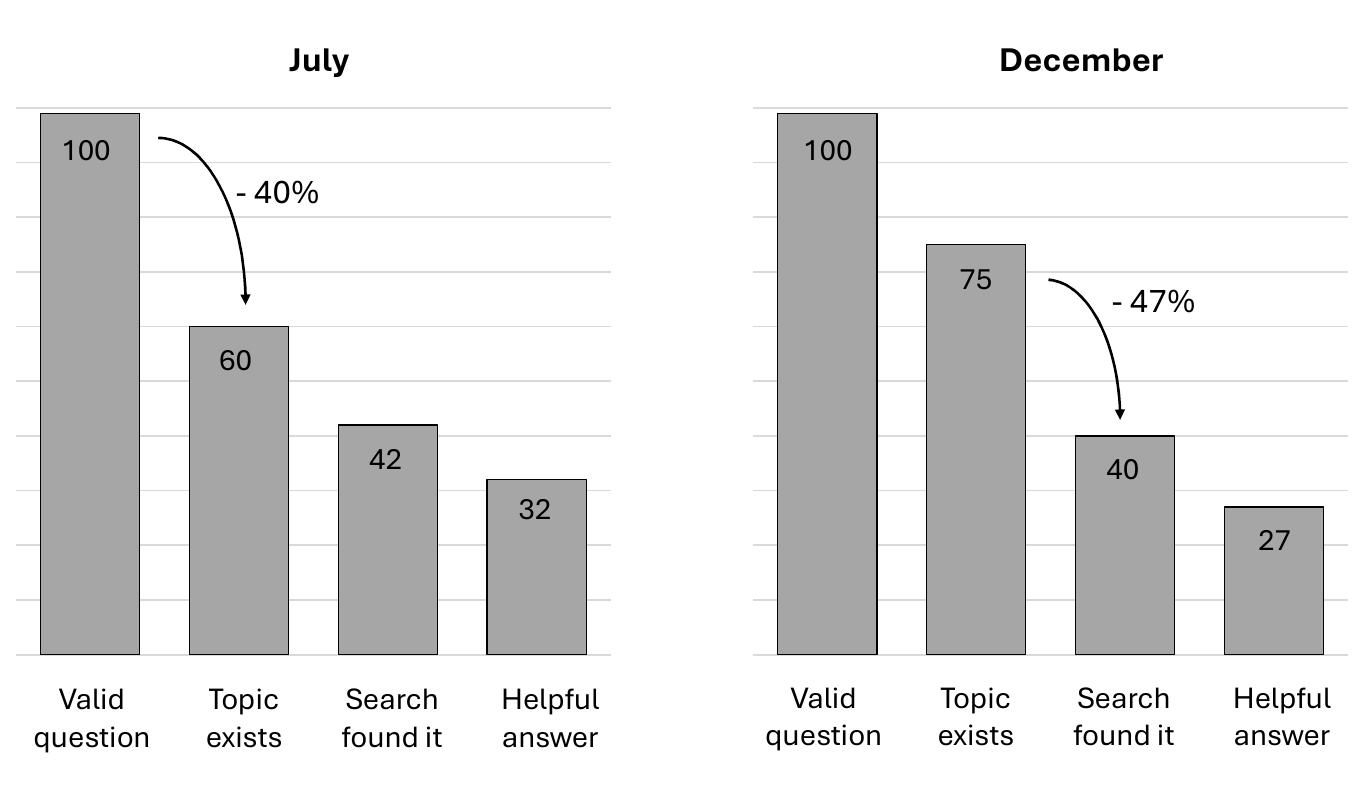}
\caption{RAG evaluation results analysis}
\Description{Bar charts for July and December.  In July, a drop of 40 percent is easy to see from the bar labelled "Valid question" to the bar labelled "Topic exists".  In the December chart, that drop is gone.  However, a drop of 47 percent is easy to see from the bar labelled "Topic exists" to the bar labelled "Search found it".}
\label{analysis}
\end{figure}

Fig. \ref{analysis} shows evaluation results for a RAG solution at two points in time.  In July, for 40\% of valid questions there were no topics containing information to answer the question.  So, we recruited writers to fill that content gap. By December, there were topics to answer valid questions 75\% of the time - an improvement. Unfortunately, search performance declined.  In December, search didn't find the relevant topic 47\% of the time.  We were able to collect sample answers evaluated as \verb|"Article exists" == "Yes"| and \verb|"Search success" == "Fail"| so we could identify and fix the cause of the search failures.  The RAG evaluation tool helps us know where to focus our improvement efforts.

\pdfbookmark[subsection]{Unanticipated benefits}{Unanticipated benefits}
\subsection*{Unanticipated benefits}

Methodical evaluation of results has increased the business value of our RAG solutions:

\begin{itemize}
\item \textbf{Training data} - Sample questions, fine-tuning data sets, custom NLP dictionaries, and training data for classifiers naturally created as we manually evaluate, annotate, and tag results can sometimes be used to improve other AI solutions, such as analysing customer feedback surveys or community questions (subject to terms of use and reuse.)
\item \textbf{Insights} - The RAG evaluation tool sends a weekly summary of user questions to a team Slack channel so everyone knows what our users are struggling with and asking about.
\item \textbf{Documentation improvements} - Our team meets for 30 minutes a week to review results and fix problems, including: content gaps, search failures, and content that needs editing.
\end{itemize}

%% file: ICAAI-2024_11_scale.tex
\section{Scaling an Enterprise solution}

When building a RAG solution to support a portfolio of dozens or hundreds of software products, new challenges arise:

\textbf{Questions vary by product} - While common questions for one product might be factual \enquote{what-is} questions, common questions for another might be command-line syntax questions.  A given question rewriting method might work for one but not the other.

\textbf{Content varies by product} - The documentation for one product might be conceptual or task-based, while another product's documentation might be mostly API reference details.  Search that works well for one might not work well for the other.

\textbf{Getting buy-in} - When one product team decides to build a RAG solution, they feel invested and prepared to do manual work like evaluating results.  But getting buy-in for an enterprise-wide initiative can be challenging.

\textbf{One size might not fit all} - Questions and content are not the only things that will vary across teams.  Building one solution for everyone maximizes shared infrastructure.  Ensuring that solution is configurable and flexible empowers individual teams to benefit from the centralized infrastructure while also doing what works best for them.

\textbf{Automated regression testing} - As teams rewrite their content and update components of their RAG solution, they need a way to test the performance of their solution without having to manually evaluate those test results.  Solutions like the RAG evaluation tool can be used to collect question-topic-answer triplets that can be tested in automated batches. (Evaluation techniques like BLEU, ROUGE, and so on, \emph{do} have a useful place here.)

%% file: ICAAI-2024_12_thanks.tex
\section*{Acknowledgement}

For supporting our work on these projects, we want to express our appreciation to our managers: Richard Horsfall, Wendy Switzer, Kirti Gani, and Lindsay Martin.  Thank you!